\documentclass[ reprint,amsmath,amssymb,aps,pre,groupedaddress]{revtex4}

\usepackage{graphicx}
\usepackage{braket}
\usepackage{dcolumn}
\usepackage{bm}

\begin{document}
\preprint{APS/123-QED}

\title{Effects of heterogeneity in site-site couplings for tight-binding models on scale-invariant structures}

\author{Bingjia Yang}
\author{Pinchen Xie}
\email{xiepc14@fudan.edu.cn}
\affiliation {Department of Physics, Fudan University,
Shanghai 200433, China}
\affiliation {Shanghai Key Lab of Intelligent Information
Processing, Fudan University, Shanghai 200433, China}

\author{Zhongzhi Zhang}
\email{zhangzz@fudan.edu.cn}
\homepage{http://www.researcherid.com/rid/G-5522-2011}
\affiliation {School of Computer Science, Fudan University,
Shanghai 200433, China}
\affiliation {Shanghai Key Lab of Intelligent Information
Processing, Fudan University, Shanghai 200433, China}

\date{\today}

\begin{abstract}
	We studied the thermodynamic behaviors of non-interacting bosons and fermions  trapped by a scale-invariant branching structure of adjustable degree of heterogeneity. The full energy spectrum in tight-binding approximation was analytically solved . We found that the log-periodic oscillation of the specific heat  for Fermi gas depended on the heterogeneity of hopping. Also, low dimensional Bose-Einstein condensation occurred only for non-homogeneous setup. 
\end{abstract}

\maketitle

\newcommand{\G}{\mathcal{G}}
\renewcommand{\S}{\mathcal{S}}
\renewcommand{\H}{\mathcal{H}}
\newcommand{\T}{\mathcal{T}}
\renewcommand{\L}{\mathcal{L}}

\section{Introduction}\label{intro}
Tight-binding   quantum gases upon quasiperiodic or fractal-like structures with scale symmetry have been studied intensively over the past few decades~\cite{Cardoso2008a, Nandy2014, Yamada2015, Jana2010, Chakrabarti2011, VanVeen2016, Moreira2008, Mauriz2001, DeOliveira2004, Coronado2005, Mauriz2003, Vallejos1998, DeOliveira2009, Aydin2014, Vidal2011, Ketterle1996, Burioni2001a, DeOliveira2013a, Bagnato1991, Buonsante2002, Serva2014a, Brunelli2004, Lyra2014a}.
 In most   cases, the energy spectrum of the ideal gas and  corresponding density of states show self-similarity and  power-law behaviors at the same time. This is responsible for a sequence of unique behaviors related to localization of states~\cite{Cardoso2008a, Nandy2014, Yamada2015},  quantum transport~\cite{Jana2010, Chakrabarti2011, VanVeen2016},  specific heat~\cite{Moreira2008, Mauriz2001, DeOliveira2004, Coronado2005, Mauriz2003, Vallejos1998, DeOliveira2009, Aydin2014},   Bose-Einstein condensation (BEC)~\cite{Vidal2011, Ketterle1996, Burioni2001a, DeOliveira2013a, Bagnato1991, Buonsante2002, Serva2014a, Brunelli2004, Lyra2014a}, etc. Though without introducing interaction, the simplest model yields lots of interesting phenomena due to the complex topology of fractal-like lattice structures. Different from isotropic models, the hopping of particles   is  non-trivial in these cases. Naturally one will ask how the heterogeneity of hopping(site-site coupling) influences the model.
 
The  heterogeneity of hopping consists of two aspects: the network topology of lattices and the variation of coupling strength.  There have been many results on how the  topology of lattice structures gives birth to unusual behaviors of hopping gases. For example, the low dimensional BEC of non-interacting bosons, trapped by diamond  hierarchical lattices,  only takes place while the branching parameter of the trap structure is lager than 2~\cite{Lyra2014a}. Recently,  the quantum transport on  Sierpinski carpets is also found to be determined  by structural parameters~\cite{VanVeen2016}. One can guess the topological properties of lattice structures decide how curved the underlying space is for the hopping gas. Though locally similar to an isotropic Euclidean lattice, a fractal-like structure can produce totally different outcomes when serving as traps for hopping gas. To describe those anisotropic structures more quantitatively, some indicators  including the fractal dimension and the spectral dimension~\cite{Cassi1993, Burioni1996, Rammal1983a, Rammal1984b, Alexander1982a} are introduced. A deterministic relation among them is also provided for some renormalizable structures~\cite{Burioni1999}.

However, it is rarely reported that how the heterogeneity of the strength of site-site couplings(hopping amplitude)  influences the behaviors of quantum gases. The heterogeneity of hopping amplitudes is worth studying since the site-site coupling is suggested to play a important role in other similar models.   There are many cases that can not be approached by   mean field approximation in real world systems. For example, the heterogeneity in the site-site coupling significantly affects the epidemic spreading ~\cite{Yang2012, Chu2011}, transportation ~\cite{Wu2013a, Soh2010}, synchronization ~\cite{Lu2006, Huang2006}, random walks~\cite{Lin2013b,Zhang2013}, diffusive processes ~\cite{Baronchelli2010},   voter models~\cite{Baronchelli2011, Suchecki2005}, etc.,  on weighted networks.  
We will fill this gap by a case study  regarding the non-interacting  Fermi  and Bose gases upon a parameterized scale-invariant  branching structure.   We will show that the heterogeneity of  coupling  strength has a  decisive influence on the thermodynamic behaviors  even in the simplest model.

This paper is organized as the following. First we construct a scale-invariant  branching structure with two parameters control the heterogeneity of our model. In tight-binding approximation we define the normalized Hamiltonian. By appropriate renormalization  the full spectrum is obtained. Then, for Fermi gas, we study its Fermi energy and subsequent log-periodic oscillation of specific heat associated to special weight parameter. As for Bose gas, we investigate its phase transition phenomenon at low temperature and find the relation between the  weight parameter and BEC.

\section{Preliminary}
\begin{figure}[b]
  \includegraphics[width=\linewidth]{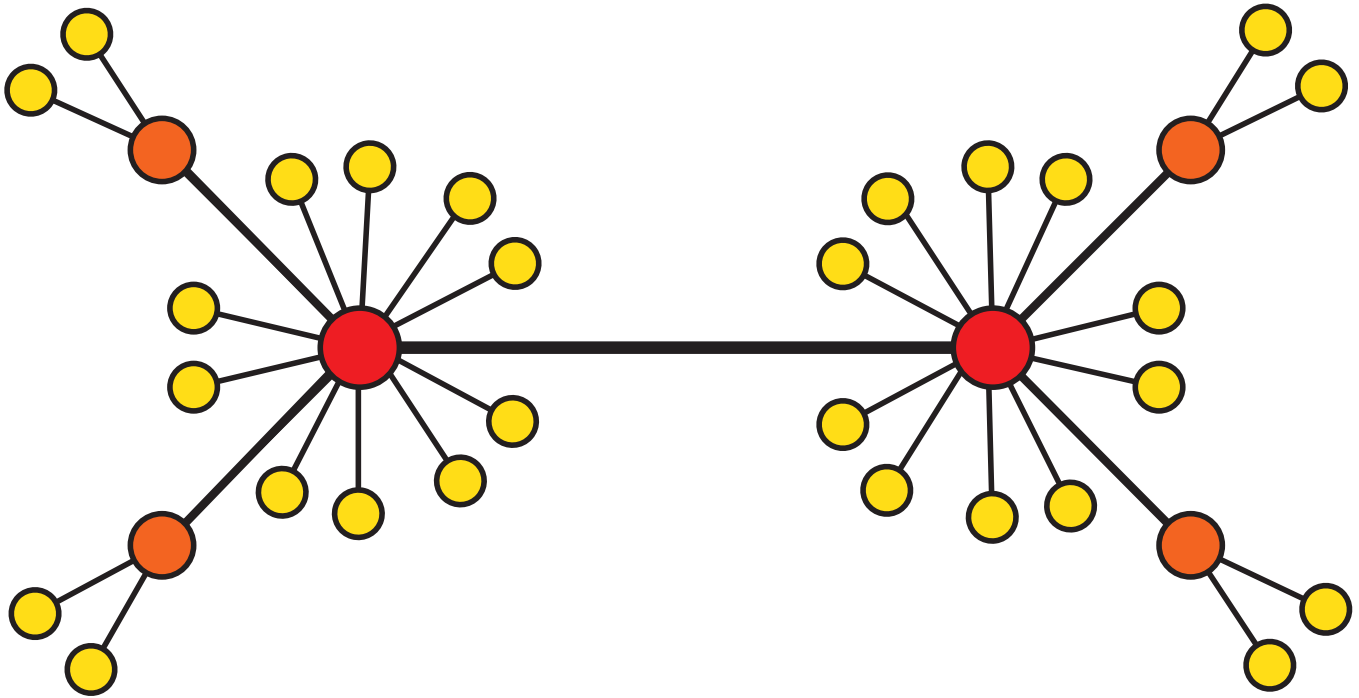}
  \caption{A sketch of G(2,1).}\label{cons}
\end{figure}

A weighted branching structure is constructed iteratively, see Fig.~\ref{cons}. 

$G^{(0)}$ is a chain of length $1$ where two vertices are connected by an edge of unit weight.
 For $t>0$, $G^{(t)}$ is obtained from $G^{(t-1)}$ by the following transformation. For each edge of weight $w$ in $G^{(t-1)}$, $mw(m>0)$ new vertices are connected to both sides of the edge respectively with unit weight, meanwhile, the weight of the old edge is increased by $m\theta w(\theta\geqslant 0 )$. The parameters $m,\theta$ are all  integers.
 Let $G(m,\theta)=\lim\limits_{t\rightarrow \infty} G^{(t)}$. An infinite branching structure forms.

 By construction, the total number of the vertices for  $G^{(t)}$ is 
 \begin{equation}
 	N_t=\frac{2}{2+\theta}[(\theta m+2m+1)^t+\theta+1].
 \end{equation}
 Name these vertices by $v_1,v_2,\cdots, v_{N_t}$. $a_{ij}$  denotes the weight of the edge connecting  $v_i$ and $v_j$. $a_{ij}$ is   $0$ when $v_i$ and $v_j$ are not adjacent. Further we define the degree of $v_i$ as $d_i=\sum_j a_{ij}$.

To describe the topological structure of $G^{(t)}$, we introduce the adjacency matrix   $(A^{(t)})_{ij}=a_{ij}$ and the degree matrix $(D^{(t)})_{ij}=\delta_{ij} d_i$. Let the normalized stochastic matrix~\cite{Chen2007} for markov chains on $G^{(t)}$ be $T^{(t)}=\sqrt{D^{(t)}}A^{(t)}\sqrt{D^{(t)}}$. Obviously, $t_{ij}=\frac{a_{ij}}{\sqrt{d_id_j}}$. For $G(m,\theta)$,  define $T=\lim\limits_{t\rightarrow\infty} T^{(t)}$.

\section{Tight-binding model on $G(m,\theta)$}\label{H}
Suppose the structure we constructed  denotes a trapping structure for quantum gases. The edges  connecting two vertices represent the correlation of two traps. 
The tight-binding Hamiltonian describing the system  writes~\cite{Serva2014a, DeOliveira2013a}



\begin{equation}\label{hamiltonian}
	\hat{\H_0}=\sum_i d_i \hat{a}_i^{\dagger }\hat{a}_i -\sum_{ij} a_{ij} \hat{a}_i^{\dagger }\hat{a}_j.
\end{equation}

Here  $\hat{a}_i^{\dagger}$ and $\hat{a}_i$ are creation and annihilation operators and    $a_{ij}$ denotes the hopping amplitude  between two coupled traps. The second summation in Eq.~(\ref{hamiltonian}) is taken over all neighboring vertices $i$ and $j$. Clearly, when $\theta=0$, $a_{ij}$ is constantly 1 for all existing site-site correlations. This is the most homogeneous case  in our model. For non-vanishing $\theta$, the hopping amplitude is heterogeneous.
 
 From Eq.~(\ref{hamiltonian}), we know the spectrum of $\hat{\H_0}$ is unbound for infinite network($t\rightarrow\infty$). However, by rescaling the frequency space(multiplying the Hamiltonian by diagonal operators at both sides),  we can  normalize $\hat{\H_0}$  as 
 
\begin{equation}\label{Ham}
	\hat{\H}=-\sum_{ij} t_{ij} \hat{a}_i^{\dagger }\hat{a}_j,
\end{equation}
of which the spectrum lies on $[-1,1]$.

The matrix $T$ we defined previously hence gives a full description of  $\hat{\H}$.  The allowed energy  for $\hat{\H}$ is  the eigenvalue spectrum of $-T$. The spectrum is  a Julia multiset $J_R\subset [-1,1]$ generated  by the inverse of the function 
\begin{equation}
	R(x)=\frac{\theta m+m+1}{\theta m+1}x-\frac{m}{(\theta m +1)x}
\end{equation}
 from $\{-1,1\}\bigcup\{0\} $~\cite{Teplyaev1998a, Hare2012}.  A detailed description of $J_R$ is provided in the appendix. Fig.~{\ref{spectrum} shows how the eigenvalue spectra vary with the weight parameter $\theta$. When $\theta\rightarrow \infty$, the spectrum is dense in $[-1, 1]$. 

Let $\textrm{deg}^{(t)}(\varepsilon)$ denote the multiplicity of the eigenvalue $\varepsilon$ of $-T^{(t)}$. Take $\textrm{deg}^{(t)}(\varepsilon)=0$ if $\varepsilon$ is not an eigenvalue. The density of states  on $[-1,1]$ is
\begin{equation}
	\rho(\varepsilon)=\sum_{\varepsilon'\in J_R}\delta(\varepsilon-\varepsilon') \lim_{t\rightarrow \infty} \frac{\textrm{deg}^{(t)}(\varepsilon' )}{N_t}
\end{equation}
where $\delta(\varepsilon-\varepsilon')$ is the Dirac delta function. Fig.~\ref{dos} is a schematic representation of $\rho(\epsilon)$ related to different $\theta$. Obviously, $\rho(\epsilon)$ shows self-similar properties.

 Fig.~\ref{spectrum} and Fig.~\ref{dos} together show that the spectrum related to $G(m,\theta)$ is highly degenerate and fractal-like. Besides,  the spectrum is  symmetric with respect to $\epsilon=0$ , which possesses the largest degeneracy.    From Fig.~\ref{spectrum}, we have also found several isolated levels, among which $\epsilon=0$ is fixed for any $\theta$. 

\begin{figure}[t]
	\includegraphics[width=\linewidth]{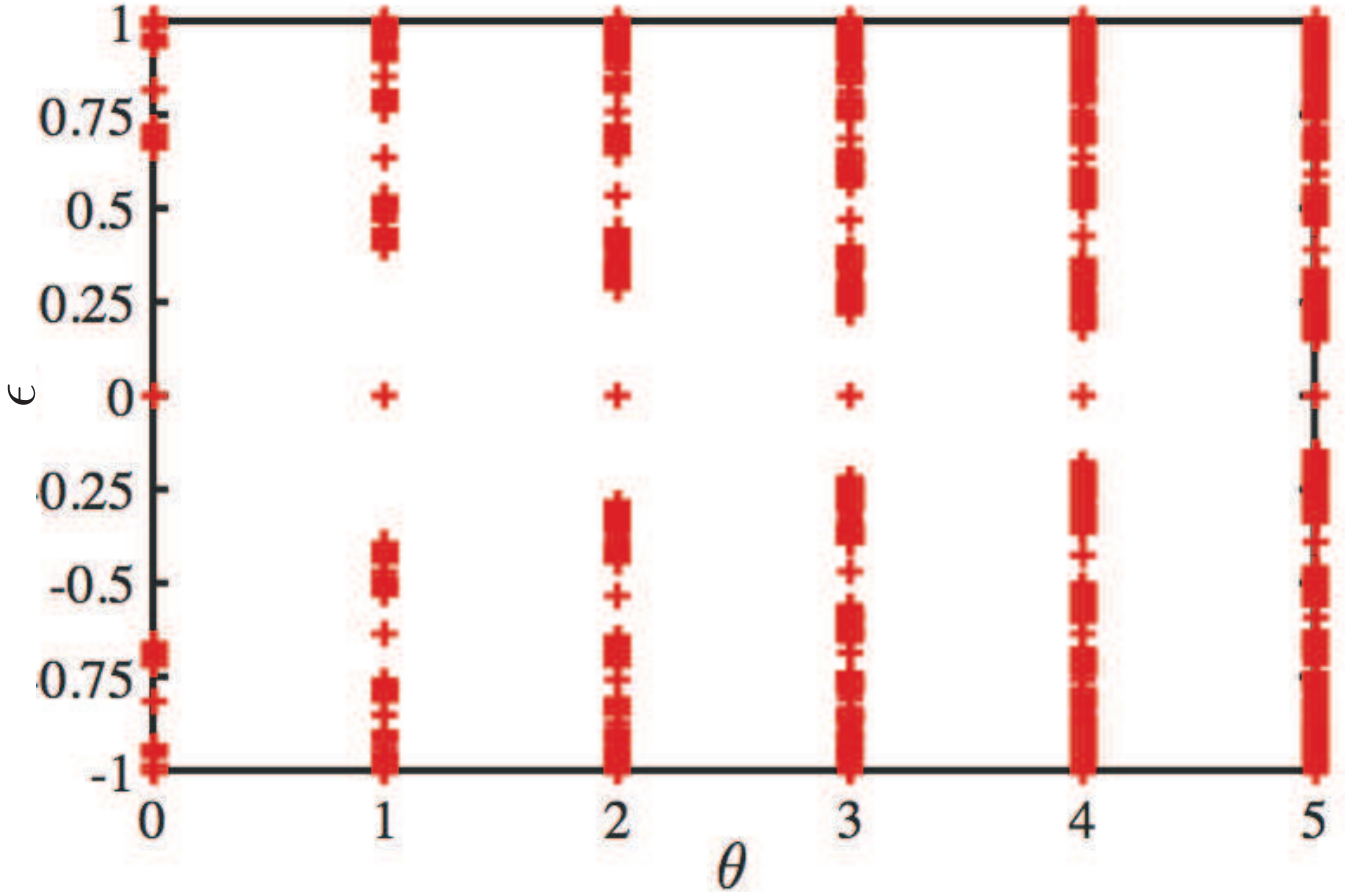}
	\caption{(Color online) Eigenvalue spectra related to  $\theta$ from $0$ to $6$ when $m=2$. }\label{spectrum}
\end{figure}

\begin{center}
\begin{figure}[t]
	\includegraphics[width=\linewidth]{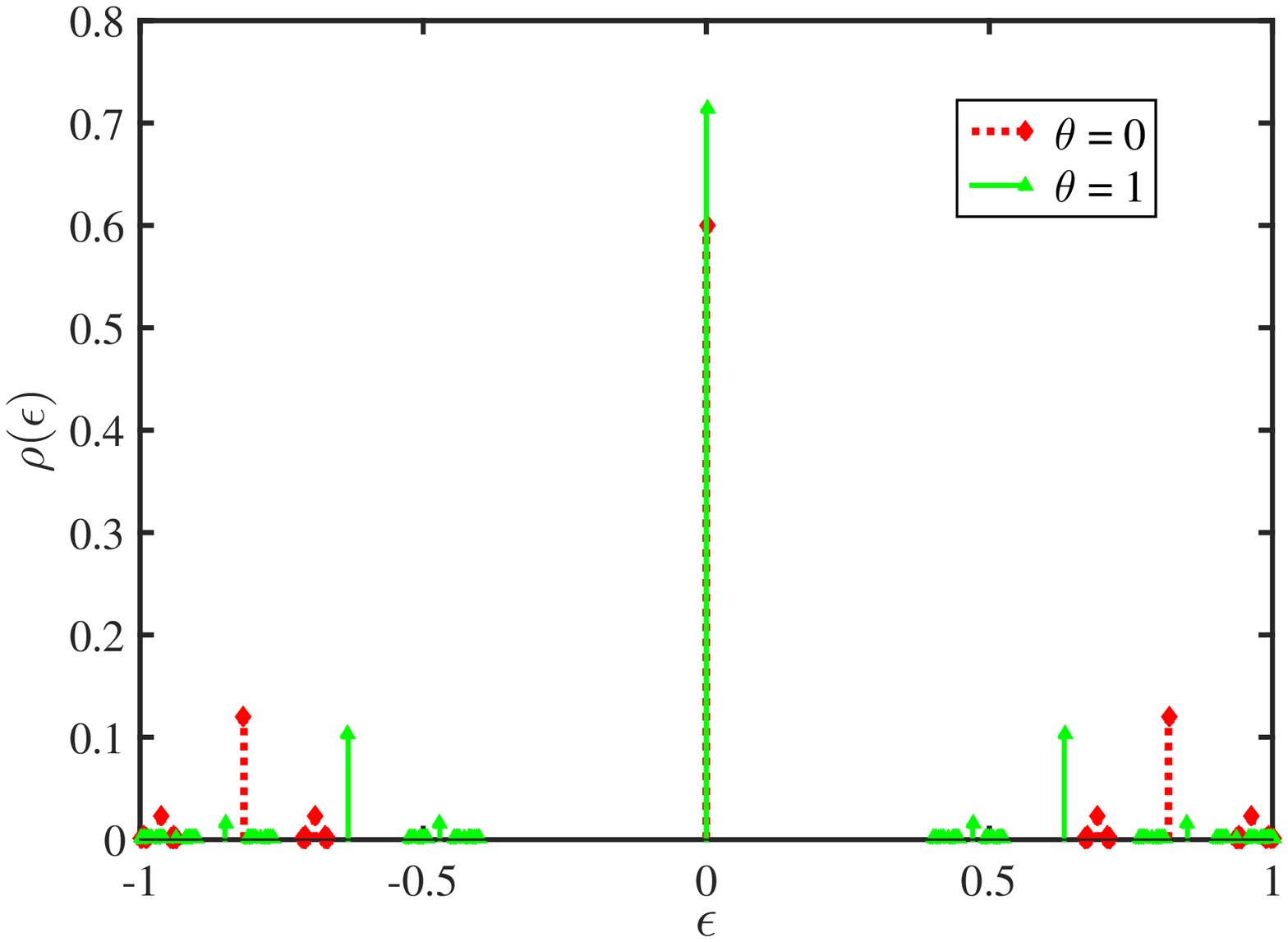}
	\caption{(Color online) Density of states associated to $G(2,0)$ and $G(2,1)$.}\label{dos}
\end{figure}
\end{center}

The integrated density of states is defined as
\begin{equation}
	I(\varepsilon)=\int_{-1}^{\varepsilon} \rho(x) dx.
\end{equation}

Since the smallest  eigenvalue of $-T$ is always $-1$ through iteration,  the lowest allowed energy for $\hat{\H}$ is $\varepsilon_0=-1$. Near the lowest energy level, the power-law behavior of the integrated density of states is observed, given by 
\begin{equation}\label{sd}
	I(\varepsilon )\sim (\varepsilon-\varepsilon_0)^{\frac{d_s}{2}}.
\end{equation}  
$d_s$ is called the  spectral dimension of $G(m,\theta)$, working as a generalization of the Euclidean dimension defined on homogeneous systems such as regular lattices~\cite{Rammal1984b, Alexander1982a,  Cassi1993}.  

Under iteration, the energy level near the bottom is stable ($R(-1)=-1$)  and the integrated density of states is fixed, one obtains
\begin{equation}\label{stable}
	\lim_{\substack{t\rightarrow \infty \\  \varepsilon \rightarrow -1}} \frac{N_{t+1}}{N_t} \frac{I(\varepsilon )}{I\left(R(\varepsilon )\right)}=1.
\end{equation}
Plugging Eq.~(\ref{sd}) into Eq.~(\ref{stable}),  after some algebra we arrive at
\begin{equation}\label{spd}
		d_s=2\frac{\ln(\theta m+2m+1)}{\ln(\theta m+2m+1)-\ln(\theta m+1)}.
\end{equation}
Obviously, $d_s\geqslant 2$, where the equality only holds for $\theta=0$. 


\section{Fermi gas on $G(m,\theta)$}
In this section we will study the behaviors of non-interacting hopping fermions trapped by the structure we have constructed. The Hamiltonian for the system is still given by Eq.~(\ref{Ham}) while the creation and annihilation operators are fermionic now.   We ignore spins for evident reasons. 

The Fermi-Dirac statistics give the average occupation number of each state with energy $\epsilon$:
\begin{equation}\label{FD}
	\bar{n}(\epsilon)=\frac{1}{e^{\beta(\epsilon-\mu)}+1}.
\end{equation}
In Eq.~(\ref{FD}), $\beta=\frac{1}{k_B T}$ and $\mu$ is the chemical potential.
The normalization condition  requires
\begin{equation}\label{norcon}
	\int_{-1}^{1} \bar{n}(\epsilon)\rho(\epsilon)d\epsilon=\gamma
\end{equation}
where $\gamma>0$ is the filling fraction, i.e., the ratio of  particles to vertices.

Let $E_F$ denote the Fermi energy of this system.
\begin{figure}[t]
	\includegraphics[width=\linewidth]{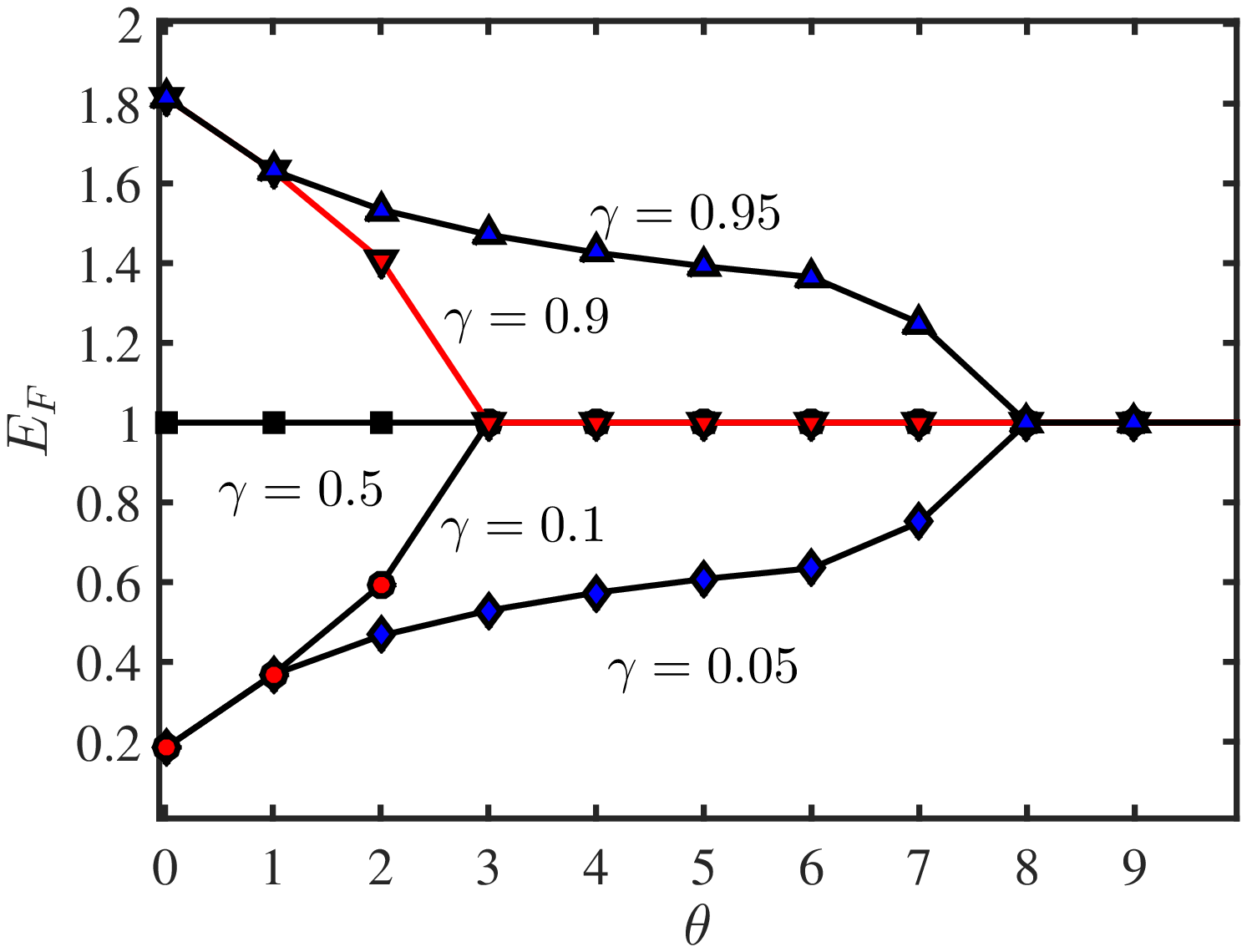}
	\caption{(Color online) The relation between $E_F$ and $\theta$  for $G(2,\theta)$.}\label{E_F}
\end{figure}
To investigate the influence of the parameter $\theta$ on the Fermi energy while the  structural parameter $m$ is fixed, we put $m=2$ without loss of generality. The dependence of $E_F$ on $\theta$ is given in Fig.~\ref{E_F}.   Except for half filling,   $E_F$ varies with $\theta$ for fixed $\gamma$ and low $\theta$ due to the huge degeneracy of  $\epsilon=0$.  We can also give the range of $\gamma$ that allows $E_F=1$ as $\gamma\in [\gamma_c(m,\theta), 1-\gamma_c(m,\theta)]$. $\gamma_c(m,\theta)$ is determined by checking the weight of  level $\epsilon=0$:
\begin{equation}\label{gammac}
	1-2\gamma_c(m,\theta)=\lim_{t\rightarrow \infty} \frac{\textrm{deg}^{(t)}(\varepsilon=0 )}{N_t}.
\end{equation}
Eq.~\ref{gammac} yields
$\gamma_c(m,\theta)=\frac{1}{\theta m+2m+1}$.
As $\theta\rightarrow\infty$, $\gamma\rightarrow 0$ since the  level $\epsilon=0$  predominates.

\begin{center}
\begin{figure}
  \includegraphics[width=\linewidth]{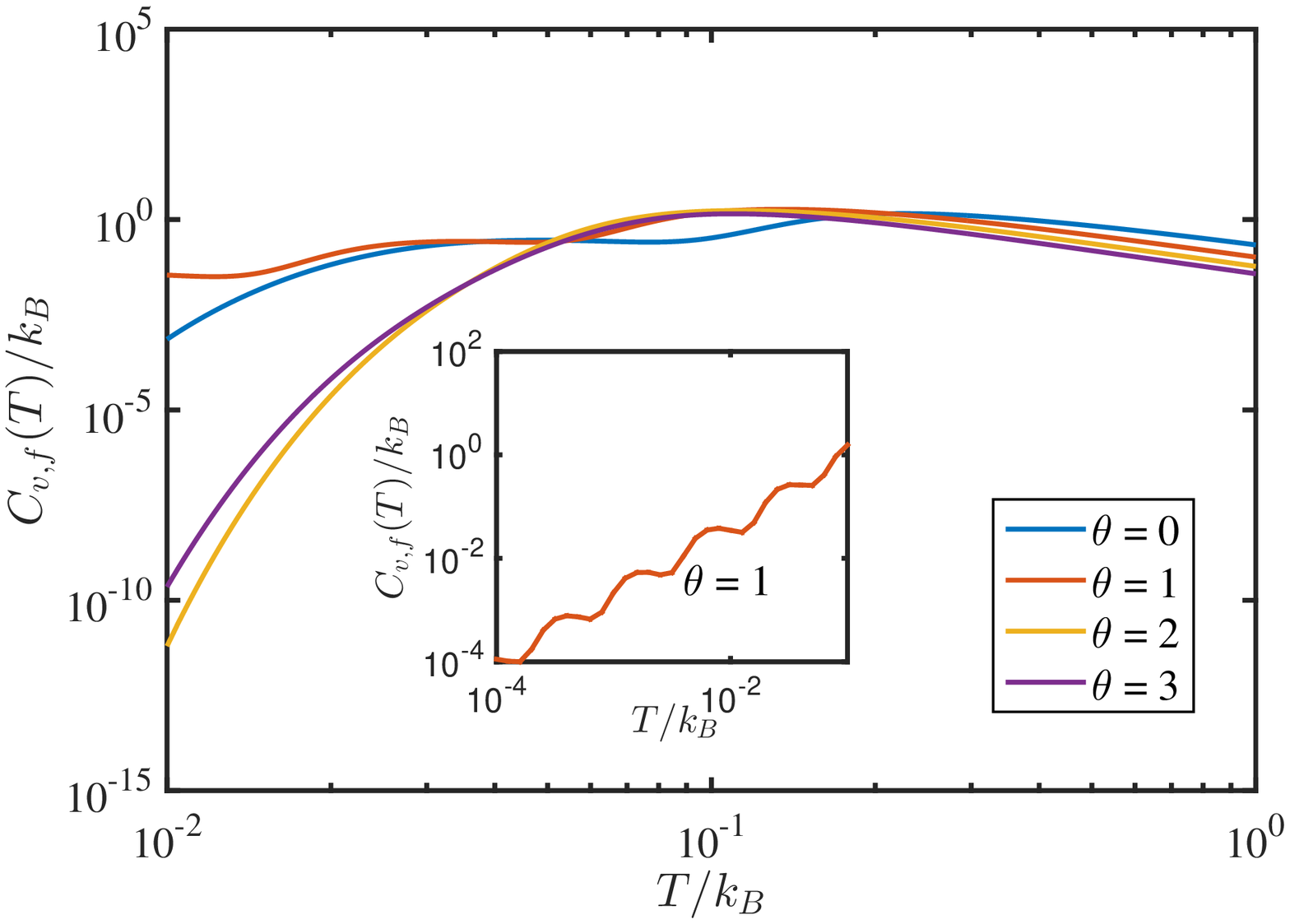}
  \caption{(Color online) Specific heat as a function of the temperature for $\theta=0,1,2,3$ ($m=2, \gamma=0.125$). The inset shows the oscillating decay of $C_{v,f}(T)$ as $T\rightarrow 0$ when $\theta=1$.}\label{fermioncapacity}
\end{figure}
\end{center}

\begin{center}
\begin{figure}
  \includegraphics[width=\linewidth]{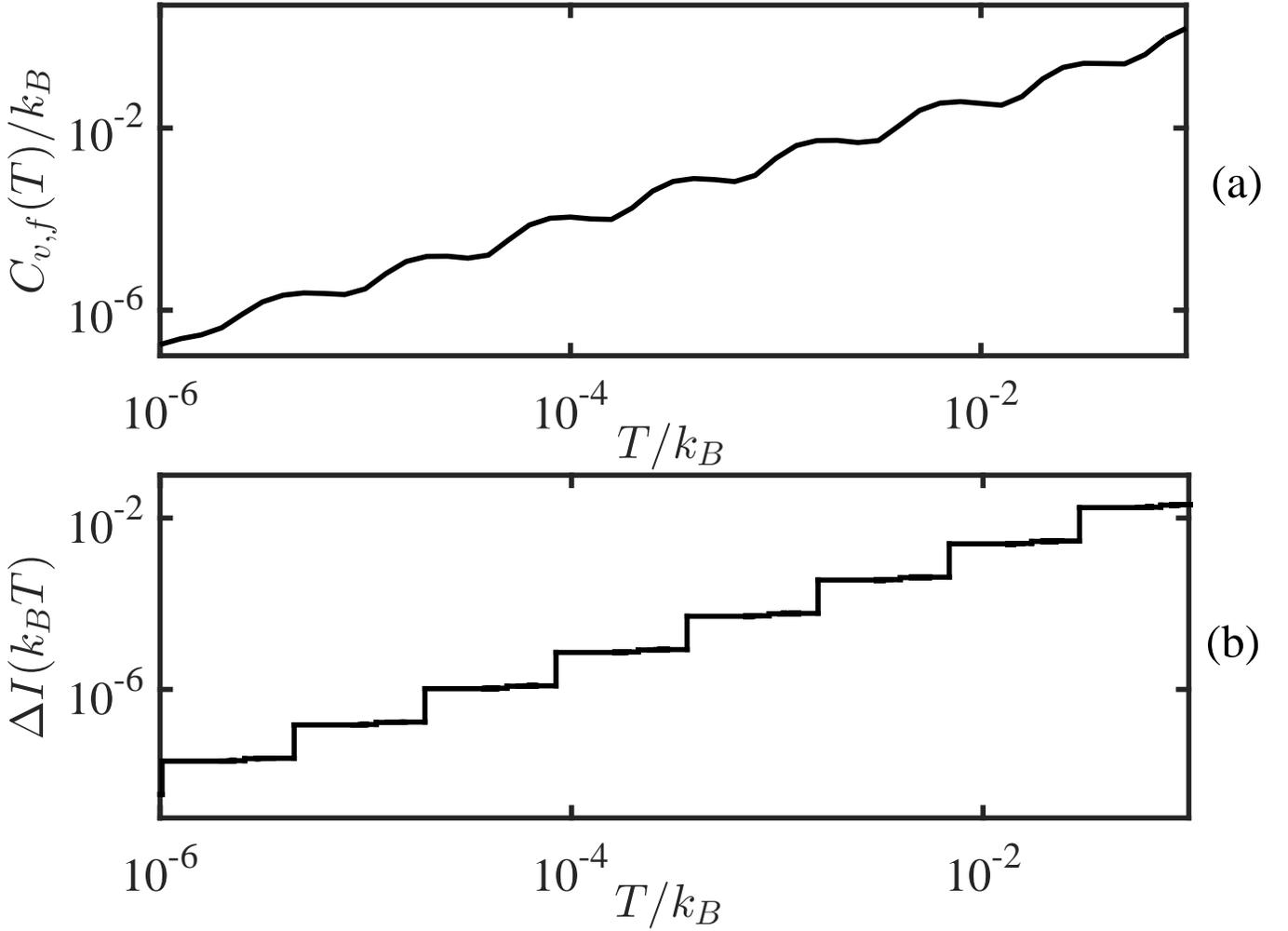}
  \caption{(a) Log-log plot of the specific heat versus the temperature ($m=2, \theta=1,\gamma=0.125$). (b) Log-log plot of $\Delta I(k_BT)=I(\epsilon_0+E_F+k_BT)-I(\epsilon_0+E_F-k_BT)$ versus the temperature ($m=2, \theta=1,\gamma=0.125$). The two curves display the same periodicity.}\label{oscillation}
\end{figure}
\end{center}

Next we pay attention to the thermodynamic property of the system.

The average internal energy per particle is 	$U=\frac{1}{\gamma}\int_{-1}^{1} \bar{n}(\epsilon)\rho(\epsilon)\epsilon d\epsilon$. Subsequently the heat capacity per particle or specific heat is 
\begin{equation}\label{capacity}
	C_{v,f}(T)=\frac{dU}{dT}=\frac{k_B\beta^2}{4}\left(<\epsilon^2\rho(\epsilon)>_{_{f}}-\frac{<\epsilon\rho(\epsilon )>_{_{f}}^2}{<\rho(\epsilon) >_{_{f}}}\right)
\end{equation}
where $<g(\epsilon)>_{_{f}}$ denotes the integral
\begin{equation}\label{bracket}
	<g(\epsilon)>_{_{f}}=\int_{-1}^1\frac{g(\epsilon)d\epsilon}{\cosh^2\frac{\beta(\epsilon-\mu)}{2}}.
\end{equation}

A schematic representation of the specific heat  as a function of temperature is given in Fig.~\ref{fermioncapacity}. 
In low-temperature approximation,   Eq.~(\ref{capacity}) becomes 
\begin{small}
\begin{equation}\label{cap2}
	C_{v,f}(T)= \frac{k_B\beta^2}{4}\left(<(\epsilon-E_F)^2\rho(\epsilon)>_{_{f}}-\frac{<(\epsilon-E_F)\rho(\epsilon )>_{_{f}}^2}{<\rho(\epsilon) >_{_{f}}}\right).
\end{equation}
The integral Eq.~(\ref{bracket}) is approximated by
\begin{equation}
	<g(\epsilon)>_{_{f}}=\int_{E_F-\delta/\beta}^{E_F+\delta/\beta}\frac{g(\epsilon)d\epsilon}{\cosh^2\frac{\beta(\epsilon-E_F)}{2}}.
\end{equation}
\end{small}
As a rough estimate we can set $1<\delta<10$.  Since $\rho(\epsilon)$ is  locally symmetric  (as shown in Fig.~\ref{dos}) and  $(\epsilon-E_F )/\cosh^2\frac{\beta(\epsilon-E_F)}{2}$ is strictly odd with respect to $\epsilon=E_F$, the second term on the right side of Eq.~(\ref{cap2}) is negligible. Thus 
\begin{equation}\label{cap3}
	C_{v,f}(T)=\frac{k_B^2T}{4}\int_{-\delta}^{\delta}\frac{t^2\rho({tk_BT}+\epsilon_0+E_F)dt}{\cosh^2\frac{ t}{2}}.
\end{equation}
If $\epsilon=\epsilon_0+E_F$ is an isolated point in the energy spectrum,  the integral in Eq.~(\ref{cap3}) is 0. Correspondingly, the curves associated to $\theta=0$ $(E_F=0.1835)$, $\theta=2$ $(E_F=1)$ and $\theta=3$ $(E_F=1)$ in Fig.~\ref{fermioncapacity} decay rapidly  as $T\rightarrow0$. On the contrary, the curve associated to   $\theta=1$ $(E_F=0.5)$ displays oscillating  decay as  $T\rightarrow 0$, because $\epsilon=\epsilon_0+E_F=-0.5$ is a limit point  in the energy spectrum. This means  $\rho(\epsilon)$ behaves scale-invariantly near $\epsilon_0+E_F$. The period of the  oscillation, thus, depends on the periodicity of $\log\int_{-\delta}^{\delta} \rho({tk_BT}+\epsilon_0+E_F)dt$ with respect to $\log T$. Fig.~\ref{oscillation} gives a schematic representation of  such relation.  One  finds $\log(C_{v,f}(T))$ has the same period as $\log\Delta I(k_BT)$ with respect to $\log(T)$ at low temperature. This unique effect   relates  the  low-temperature dependence of the specific heat to the self-similarity of the spectrum  straightforwardly. Moreover, this effect is only present for certain degree of heterogeneity of the coupling for fixed $\gamma$.  And, in large $\theta$, this effect is only allowed for full filling.  Other results regarding the log-periodic oscillation of specific heat can be found at~\cite{DeOliveira2009, Aydin2014}.

Based on above analysis, the value of the structural parameters $m$, $\theta$ and the filling fraction $\gamma$ together determine the  low-temperature dependence of the specific heat. The trivial case occurs when  $\gamma\in [\gamma_c(m,\theta), 1-\gamma_c(m,\theta)]$, for which  $E_F=1$ is a isolated energy. Otherwise we need to check whether  $E_F$ represents an non-isolated level so that the specific heat will display log-periodic oscillation  at low temperature.

\section{Bose gas on $G(m,\theta)$}
As for non-interacting hopping bosons on the same structures, we focus on the low-dimensional Bose-Einstein condensation(BEC) induced by the heterogeneity of site-site coupling. BEC purely arising from  topological  heterogeneity  has been studied before, see~\cite{Lyra2014a}. So here we still keep $m=2$.

Ignoring spin, the average occupation number of each state is  given by  
\begin{equation}
	\bar{n}(\epsilon)=\frac{1}{e^{\beta(\epsilon-\mu)}-1}.
\end{equation}
At low temperature,  	$n_0(T)=\frac{1}{e^{\beta(\varepsilon_0-\mu )}-1}$
is the number of condensed bosons and $f_0(T)=n_0(T)/N_t$ the condensed fraction.

The normalization condition is still Eq.~(\ref{norcon}).
And the specific heat is 
\begin{equation}
	C_{v,b}=\frac{k_B\beta^2}{4}\left(<\epsilon^2\rho(\epsilon )>_{_{b}}-\frac{<\epsilon\rho(\epsilon )>_{_{b}}^2}{< \rho(\epsilon) >_{_{b}}}\right)
\end{equation}
where $<g(\epsilon)>_{_{f}}$ denotes the integral
\begin{equation}
	<g(\epsilon)>_{_{b}}=\int_{-1}^1\frac{g(\epsilon)d\epsilon}{\sinh^2\frac{\beta(\epsilon-\mu)}{2}}.
\end{equation}

\begin{center}
	
 \begin{figure}[t]
  \includegraphics[width=\linewidth]{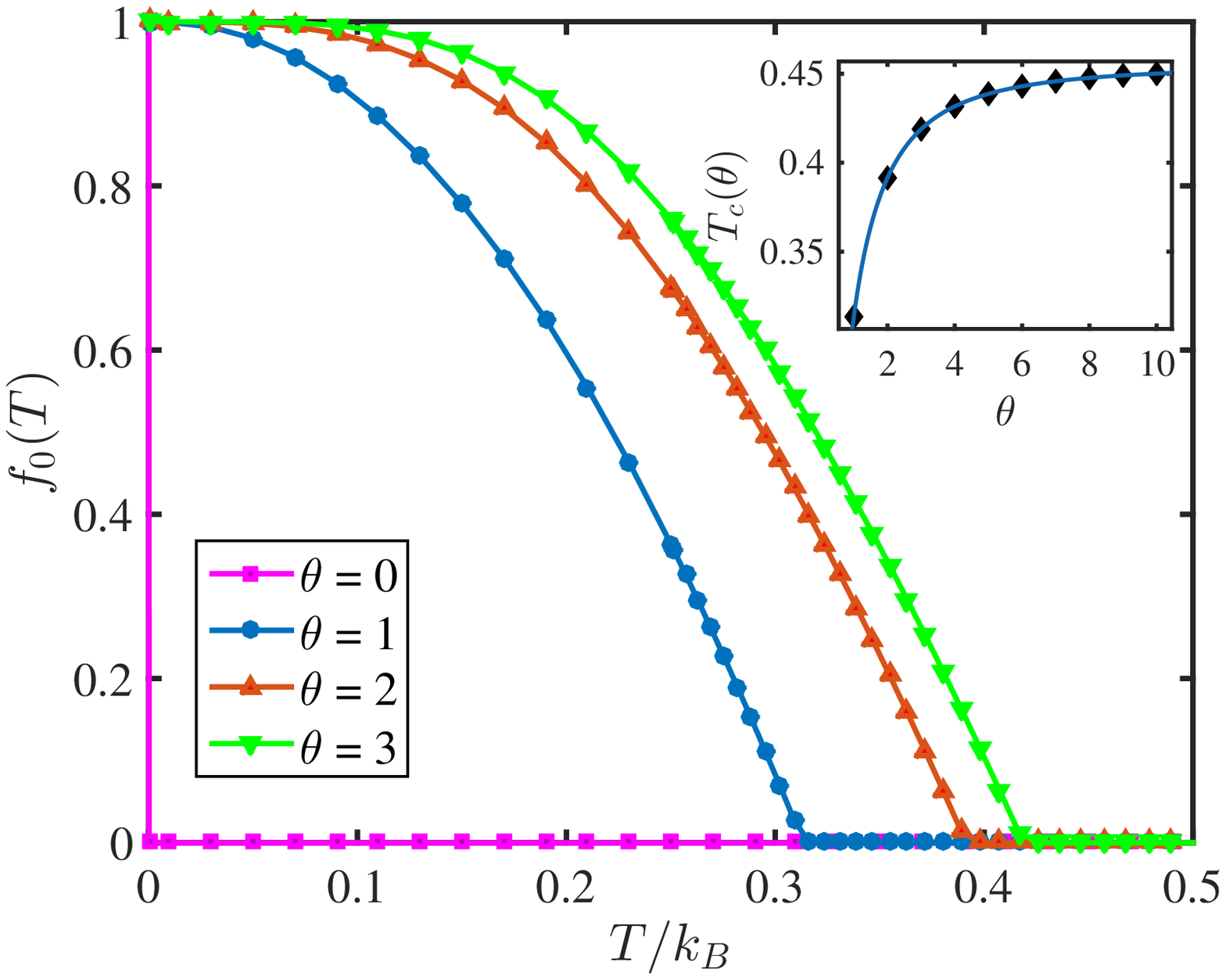}
  \caption{(Color online) Condensed fraction as a function of temperature for $\theta=0$ $ (d_s=2)$, $1$ $(d_s=4.593)$, $2$ $(d_s=7.476)$, $3$ $(d_s=10.610)$ when the filling fraction is $\gamma=1/8$. For $\theta>0$, Bose-Einstein condensation is detected. The inset displays the transition temperature as a function of $\theta$ for $\gamma=1/8$. It is saturated as $\theta\rightarrow \infty $. }\label{condensed}
\end{figure}
\end{center}
\begin{center}
	
\begin{figure}[t]
  \includegraphics[width=\linewidth]{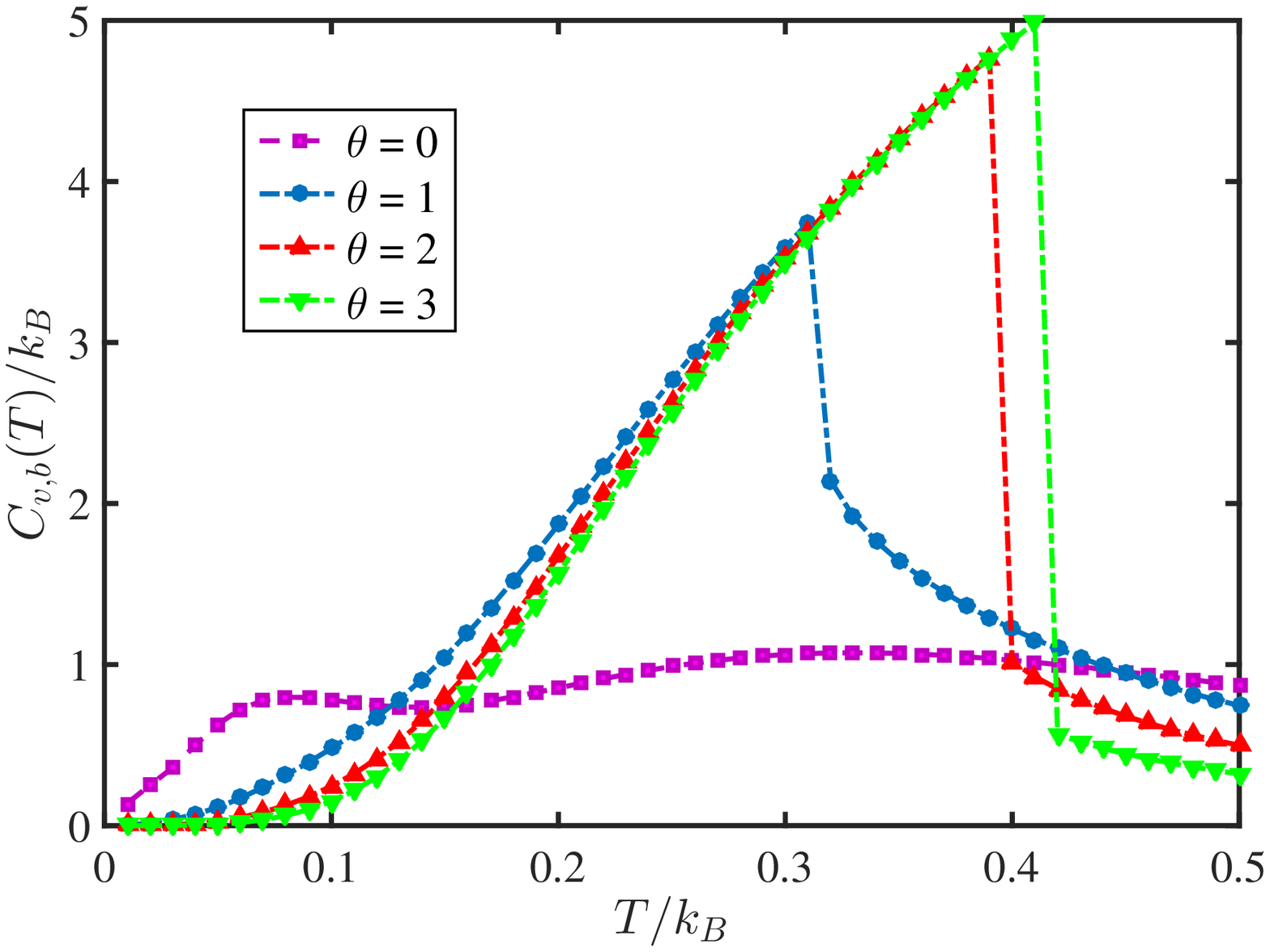}
  \caption{(Color online) Specific as a function of temperature $\theta=0,1,2,3$ when $\gamma=1/8$. $C_{v,b}(T)$ has jump discontinuity at the transition temperature for $\theta>0$. }\label{bosoncapacity}
\end{figure}
\end{center}

A schematic representation of the relation between the condensed fraction  and the temperature  is given by Fig.~\ref{condensed}, where we find the value of $\theta$ determines the occurrence of BEC and the transition temperature.  The corresponding relation between the specific heat and the temperature is shown in Fig.~\ref{bosoncapacity}. It is clear that BEC occurs only for non-vanishing $\theta$. In Fig.~\ref{bosoncapacity} we also observe  that the curves associated to $\theta>0$ have jump discontinuities, indicating the location of the transition temperature. Obviously, the transition temperature $T_c(\theta)$  increases monotonically and saturates when $\theta \rightarrow \infty$, as shown in the inset of Fig.~\ref{condensed}. As one can predict, the critical relation for condensed fraction is $f_0(T)\propto\frac{T_c-T}{Tc}$ near the critical temperature.

Through Eq.~(\ref{spd}) we find the spectral dimension  related to $\theta=0$ is always $2$ no matter how large $m$ is. Since  BEC is forbidden for spectral dimension not larger than 2, BEC does not take place upon $G(m,0)$ that is the homogeneous configuration. This conclusion exactly reflects the importance of the heterogeneity of the site-site coupling strength instead of the heterogeneity in topological structures. Because the latter fails to induce a phase transition.

\section{Conclusion}
Though it may be satisfactory to study homogeneous model for many real-world solids when  disorder is introduced, the non-uniform model is necessary for plenty of  other applications as we have mentioned in Sec.~\ref{intro}. 
In this spirit, we investigated trapped non-interacting bosons and fermions upon a scale-invariant  trap structure, with the  branching parameter $m$ and the weight parameter $\theta$ controlling the heterogeneity of the site-site coupling strength. 
Through exact renormalization, we  analytically determined the entire energy spectrum of the normalized Hamiltonian with tight-binding approximation.  The spectrum was shaped by the value of $\theta$.
Then we studied the unique thermodynamic behaviors of our model at low temperature with different degree of heterogeneity.
 To fermions, log-periodic oscillation of specific heat  only occurs for certain $\theta$ provided fixed filling fraction.  To bosons, the Bose-Einstein condensation is forbidden for the homogeneous setup($\theta=0$) regardless of the value of $m$. 

The scale-invariant branching structure in this study is exactly renormalizable hence allows efficient analytical study.  At the same time other irregular structures can not enjoy such convenience. Nevertheless, the deterministic renormalizable structures give us precious insight into the heterogeneity of network-like structures. Be more specific, our model serves as a good prototype for plenty of scale-invariant models  all displaying fractal-like spectra~\cite{Cardoso2008a, Nandy2014}.
In fact one can distort a solved structure while keeping the emergent behaviors unchanged. For example, the branching parameter $m$ in our model can  actually vary with different branches while keeping BEC forbidden for $\theta=0$. 

As a final comment, if we regard the topological aspect of structural heterogeneity as the shape of a circuit, the heterogeneity in site-site couplings alters the distribution of resistance. The latter puts the emphasis on some parts of a lattice and weakens the rest. In this circumstance, more  features are added to a model that is previously uniform. This is the reason why we can find lots of interesting phenomena in the simplest model we consider here. 
\begin{acknowledgements}
	This work was supported by the National Natural Science Foundation of China under Grant No. 11275049. P. Xie and B. Yang were also supported by Fudan’s Undergraduate Research Opportunities Program, FDUROP. Also, P. Xie appreciated the fruitful discussion with Jin-Yue Su. 
\end{acknowledgements}

\section*{Appendix} 
For clarity, name  the vertex set of $G^{(t)}$ by $V_t$. The set of the new nodes generated at iteration $t+1$ is denoted by $V_{t+1} \backslash V_{t} $.
By proper arranging   $T^{(t+1)}$ can be expressed as
\begin{equation}
	T^{(t+1)}=\left(\begin{matrix}\label{a1}
  \frac{\theta m+1}{\theta m+m+1}T^{(t)}& S \\
 S^{T}& 0
\end{matrix}
	\right).
\end{equation}
In Eq.~(\ref{a1}) $S$ reflects the  coupling between the  new vertices generated at iteration $t+1$ and the rest. Correspondingly partition the eigenvector of $T^{(t+1)}$ as $\psi=(\phi , \varphi)$ with respect to eigenvalue $\lambda$. Hence
	\begin{alignat}{2}
		\frac{\theta m+1}{\theta m+m+1}T^{(t)}&\phi+S\varphi &=\lambda\phi  \\
		S^{T}&\phi &=\lambda \varphi,
	\end{alignat}
which yields
\begin{equation}\label{a2}
\bigg( \frac{\lambda(\theta m+1)}{\theta m+m+1}T^{(t)}+SS^{T}-\lambda^2 \bigg)\phi=0
\end{equation}
provided $\lambda\neq 0$.

By the construction procedure, 
\begin{equation}
	(SS^{T})_{ij}=\sum_{v_k\in V_{t+1}\backslash V_t} S_{ik}S_{jk}=\sum_{v_k\in V_{t+1}\backslash V_t} \frac{a_{ik}a_{jk}}{d_k\sqrt{d_id_j}}
\end{equation}
is  nonzero if and only if $i=j$ since each $v_k$ is adjacent to one vertex only.
By simple algebra one arrives at
\begin{equation}\label{ss}
	(SS^{T})_{ii}=\frac{m}{\theta m+m+1}.
\end{equation}
Due to  Eq.~(\ref{ss}), Eq.~(\ref{a2}) is simplified as
\begin{equation}
\left(T^{(t)}-\frac{\theta m+m+1}{\theta m+1}\lambda+\frac{m}{(\theta m +1)\lambda}\right)\phi=0
\end{equation}
which entails
\begin{equation}
	R(\lambda)=\frac{\theta m+m+1}{\theta m+1}\lambda-\frac{m}{(\theta m +1)\lambda}.
\end{equation}
 $ R(\lambda)$ is  the eigenvalue of $T^{(t)}$ associated to eigenvector $\phi$. Further one will find the multiplicity of $R(\lambda)$ to $T^{(t)}$ is the same as the multiplicity of $\lambda$ to $T^{(t+1)}$.  
  
  Without proof  we claim that the preimage of $\lambda$ under function $R$ is a subset of the eigenvalue spectrum of $T^{(t+1)}$.  Any eigenvalue in the preimage of $\lambda$ carries on the multiplicity of $\lambda$. Notice that by tracking the preimage of the spectrum $T^{(t)}$ under function $R$ one can  obtain only $2N_t$ eigenvalues of $T^{(t+1)}$. Fortunately, the left $N_{t+1}-2N_t$ eigenvalues are uniformly 0 that is the only singularity of  $R$. More detailed discussion about the iterative derivation of spectrum can be found at~\cite{Xie2016a, Xie2015}.

So far we are able to obtain the full spectrum of $T^{(t)}$ by tracking the flow generated by $R$. As $t\rightarrow \infty$, the spectrum grows into a Julia set $J_R$ given as 
\begin{equation}
	J_R=\{0\} \bigcup \lim_{n\rightarrow\infty}R^{-n}(\{1,-1,0\}).
\end{equation}

As for $-T$, the result is the same since the spectrum of $T$ is symmetric about $0$.

\bibliographystyle{apsrev}
\bibliography{tb.bib}

\begin{thebibliography}{46}%
\makeatletter
\providecommand \@ifxundefined [1]{%
 \@ifx{#1\undefined}
}%
\providecommand \@ifnum [1]{%
 \ifnum #1\expandafter \@firstoftwo
 \else \expandafter \@secondoftwo
 \fi
}%
\providecommand \@ifx [1]{%
 \ifx #1\expandafter \@firstoftwo
 \else \expandafter \@secondoftwo
 \fi
}%
\providecommand \natexlab [1]{#1}%
\providecommand \enquote  [1]{``#1''}%
\providecommand \bibnamefont  [1]{#1}%
\providecommand \bibfnamefont [1]{#1}%
\providecommand \citenamefont [1]{#1}%
\providecommand \href@noop [0]{\@secondoftwo}%
\providecommand \href [0]{\begingroup \@sanitize@url \@href}%
\providecommand \@href[1]{\@@startlink{#1}\@@href}%
\providecommand \@@href[1]{\endgroup#1\@@endlink}%
\providecommand \@sanitize@url [0]{\catcode `\\12\catcode `\$12\catcode
  `\&12\catcode `\#12\catcode `\^12\catcode `\_12\catcode `\%12\relax}%
\providecommand \@@startlink[1]{}%
\providecommand \@@endlink[0]{}%
\providecommand \url  [0]{\begingroup\@sanitize@url \@url }%
\providecommand \@url [1]{\endgroup\@href {#1}{\urlprefix }}%
\providecommand \urlprefix  [0]{URL }%
\providecommand \Eprint [0]{\href }%
\providecommand \doibase [0]{http://dx.doi.org/}%
\providecommand \selectlanguage [0]{\@gobble}%
\providecommand \bibinfo  [0]{\@secondoftwo}%
\providecommand \bibfield  [0]{\@secondoftwo}%
\providecommand \translation [1]{[#1]}%
\providecommand \BibitemOpen [0]{}%
\providecommand \bibitemStop [0]{}%
\providecommand \bibitemNoStop [0]{.\EOS\space}%
\providecommand \EOS [0]{\spacefactor3000\relax}%
\providecommand \BibitemShut  [1]{\csname bibitem#1\endcsname}%
\let\auto@bib@innerbib\@empty
\bibitem [{\citenamefont {Cardoso}\ \emph {et~al.}(2008)\citenamefont
  {Cardoso}, \citenamefont {Andrade},\ and\ \citenamefont
  {Souza}}]{Cardoso2008a}%
  \BibitemOpen
  \bibfield  {author} {\bibinfo {author} {\bibfnamefont {A.~L.}\ \bibnamefont
  {Cardoso}}, \bibinfo {author} {\bibfnamefont {R.~F.~S.}\ \bibnamefont
  {Andrade}}, \ and\ \bibinfo {author} {\bibfnamefont {A.~M.~C.}\ \bibnamefont
  {Souza}},\ }\href {\doibase 10.1103/PhysRevB.78.214202} {\bibfield  {journal}
  {\bibinfo  {journal} {Phys. Rev. B}\ }\textbf {\bibinfo {volume} {78}},\
  \bibinfo {pages} {214202} (\bibinfo {year} {2008})}\BibitemShut {NoStop}%
\bibitem [{\citenamefont {Nandy}\ \emph {et~al.}(2014)\citenamefont {Nandy},
  \citenamefont {Pal},\ and\ \citenamefont {Chakrabarti}}]{Nandy2014}%
  \BibitemOpen
  \bibfield  {author} {\bibinfo {author} {\bibfnamefont {A.}~\bibnamefont
  {Nandy}}, \bibinfo {author} {\bibfnamefont {B.}~\bibnamefont {Pal}}, \ and\
  \bibinfo {author} {\bibfnamefont {A.}~\bibnamefont {Chakrabarti}},\ }\href
  {\doibase 10.1016/j.physleta.2014.09.012} {\bibfield  {journal} {\bibinfo
  {journal} {Phys. Lett. A}\ }\textbf {\bibinfo {volume} {378}},\ \bibinfo
  {pages} {3144} (\bibinfo {year} {2014})}\BibitemShut {NoStop}%
\bibitem [{\citenamefont {Yamada}(2015)}]{Yamada2015}%
  \BibitemOpen
  \bibfield  {author} {\bibinfo {author} {\bibfnamefont {H.~S.}\ \bibnamefont
  {Yamada}},\ }\href {\doibase 10.1140/epjb/e2015-60363-3} {\bibfield
  {journal} {\bibinfo  {journal} {Eur. Phys. J. B}\ }\textbf {\bibinfo {volume}
  {88}},\ \bibinfo {pages} {264} (\bibinfo {year} {2015})}\BibitemShut
  {NoStop}%
\bibitem [{\citenamefont {Jana}\ \emph {et~al.}(2010)\citenamefont {Jana},
  \citenamefont {Chakrabarti},\ and\ \citenamefont {Chattopadhyay}}]{Jana2010}%
  \BibitemOpen
  \bibfield  {author} {\bibinfo {author} {\bibfnamefont {S.}~\bibnamefont
  {Jana}}, \bibinfo {author} {\bibfnamefont {A.}~\bibnamefont {Chakrabarti}}, \
  and\ \bibinfo {author} {\bibfnamefont {S.}~\bibnamefont {Chattopadhyay}},\
  }\href {\doibase 10.1016/j.physb.2010.05.077} {\bibfield  {journal} {\bibinfo
   {journal} {Phys. B Condens. Matter}\ }\textbf {\bibinfo {volume} {405}},\
  \bibinfo {pages} {3735} (\bibinfo {year} {2010})}\BibitemShut {NoStop}%
\bibitem [{\citenamefont {Chakrabarti}(2011)}]{Chakrabarti2011}%
  \BibitemOpen
  \bibfield  {author} {\bibinfo {author} {\bibfnamefont {A.}~\bibnamefont
  {Chakrabarti}},\ }\href {\doibase 10.1016/j.physleta.2011.07.061} {\bibfield
  {journal} {\bibinfo  {journal} {Phys. Lett. A}\ }\textbf {\bibinfo {volume}
  {375}},\ \bibinfo {pages} {3899} (\bibinfo {year} {2011})}\BibitemShut
  {NoStop}%
\bibitem [{\citenamefont {van Veen}\ \emph {et~al.}(2016)\citenamefont {van
  Veen}, \citenamefont {Yuan}, \citenamefont {Katsnelson}, \citenamefont
  {Polini},\ and\ \citenamefont {Tomadin}}]{VanVeen2016}%
  \BibitemOpen
  \bibfield  {author} {\bibinfo {author} {\bibfnamefont {E.}~\bibnamefont {van
  Veen}}, \bibinfo {author} {\bibfnamefont {S.}~\bibnamefont {Yuan}}, \bibinfo
  {author} {\bibfnamefont {M.~I.}\ \bibnamefont {Katsnelson}}, \bibinfo
  {author} {\bibfnamefont {M.}~\bibnamefont {Polini}}, \ and\ \bibinfo {author}
  {\bibfnamefont {A.}~\bibnamefont {Tomadin}},\ }\href {\doibase
  10.1103/PhysRevB.93.115428} {\bibfield  {journal} {\bibinfo  {journal} {Phys.
  Rev. B}\ }\textbf {\bibinfo {volume} {93}},\ \bibinfo {pages} {115428}
  (\bibinfo {year} {2016})}\BibitemShut {NoStop}%
\bibitem [{\citenamefont {Moreira}\ \emph {et~al.}(2008)\citenamefont
  {Moreira}, \citenamefont {Albuquerque}, \citenamefont {da~Silva},\ and\
  \citenamefont {Galv{{a}}o}}]{Moreira2008}%
  \BibitemOpen
  \bibfield  {author} {\bibinfo {author} {\bibfnamefont {D.}~\bibnamefont
  {Moreira}}, \bibinfo {author} {\bibfnamefont {E.}~\bibnamefont
  {Albuquerque}}, \bibinfo {author} {\bibfnamefont {L.}~\bibnamefont
  {da~Silva}}, \ and\ \bibinfo {author} {\bibfnamefont {D.}~\bibnamefont
  {Galv{{a}}o}},\ }\href {\doibase 10.1016/j.physa.2008.06.004} {\bibfield
  {journal} {\bibinfo  {journal} {Phys. A Stat. Mech. its Appl.}\ }\textbf
  {\bibinfo {volume} {387}},\ \bibinfo {pages} {5477} (\bibinfo {year}
  {2008})}\BibitemShut {NoStop}%
\bibitem [{\citenamefont {Mauriz}\ \emph {et~al.}(2001)\citenamefont {Mauriz},
  \citenamefont {Albuquerque},\ and\ \citenamefont {Vasconcelos}}]{Mauriz2001}%
  \BibitemOpen
  \bibfield  {author} {\bibinfo {author} {\bibfnamefont {P.}~\bibnamefont
  {Mauriz}}, \bibinfo {author} {\bibfnamefont {E.}~\bibnamefont {Albuquerque}},
  \ and\ \bibinfo {author} {\bibfnamefont {M.}~\bibnamefont {Vasconcelos}},\
  }\href {\doibase 10.1016/S0378-4371(01)00021-8} {\bibfield  {journal}
  {\bibinfo  {journal} {Phys. A Stat. Mech. its Appl.}\ }\textbf {\bibinfo
  {volume} {294}},\ \bibinfo {pages} {403} (\bibinfo {year}
  {2001})}\BibitemShut {NoStop}%
\bibitem [{\citenamefont {de~Oliveira}\ \emph {et~al.}(2004)\citenamefont
  {de~Oliveira}, \citenamefont {Lyra},\ and\ \citenamefont
  {Albuquerque}}]{DeOliveira2004}%
  \BibitemOpen
  \bibfield  {author} {\bibinfo {author} {\bibfnamefont {I.}~\bibnamefont
  {de~Oliveira}}, \bibinfo {author} {\bibfnamefont {M.}~\bibnamefont {Lyra}}, \
  and\ \bibinfo {author} {\bibfnamefont {E.}~\bibnamefont {Albuquerque}},\
  }\href {\doibase 10.1016/j.physa.2004.05.059} {\bibfield  {journal} {\bibinfo
   {journal} {Phys. A Stat. Mech. its Appl.}\ }\textbf {\bibinfo {volume}
  {343}},\ \bibinfo {pages} {424} (\bibinfo {year} {2004})}\BibitemShut
  {NoStop}%
\bibitem [{\citenamefont {Coronado}\ and\ \citenamefont
  {Carpena}(2005)}]{Coronado2005}%
  \BibitemOpen
  \bibfield  {author} {\bibinfo {author} {\bibfnamefont {A.~V.}\ \bibnamefont
  {Coronado}}\ and\ \bibinfo {author} {\bibfnamefont {P.}~\bibnamefont
  {Carpena}},\ }\href {\doibase 10.1016/j.physa.2005.04.018} {\bibfield
  {journal} {\bibinfo  {journal} {Phys. A Stat. Mech. its Appl.}\ }\textbf
  {\bibinfo {volume} {358}},\ \bibinfo {pages} {299} (\bibinfo {year}
  {2005})}\BibitemShut {NoStop}%
\bibitem [{\citenamefont {Mauriz}\ \emph {et~al.}(2003)\citenamefont {Mauriz},
  \citenamefont {Vasconcelos},\ and\ \citenamefont {Albuquerque}}]{Mauriz2003}%
  \BibitemOpen
  \bibfield  {author} {\bibinfo {author} {\bibfnamefont {P.~W.}\ \bibnamefont
  {Mauriz}}, \bibinfo {author} {\bibfnamefont {M.~S.}\ \bibnamefont
  {Vasconcelos}}, \ and\ \bibinfo {author} {\bibfnamefont {E.~L.}\ \bibnamefont
  {Albuquerque}},\ }\href {\doibase 10.1016/S0378-4371(03)00605-8} {\bibfield
  {journal} {\bibinfo  {journal} {Phys. A Stat. Mech. its Appl.}\ }\textbf
  {\bibinfo {volume} {329}},\ \bibinfo {pages} {101} (\bibinfo {year}
  {2003})}\BibitemShut {NoStop}%
\bibitem [{\citenamefont {Vallejos}\ and\ \citenamefont
  {Anteneodo}(1998)}]{Vallejos1998}%
  \BibitemOpen
  \bibfield  {author} {\bibinfo {author} {\bibfnamefont {R.~O.}\ \bibnamefont
  {Vallejos}}\ and\ \bibinfo {author} {\bibfnamefont {C.}~\bibnamefont
  {Anteneodo}},\ }\href {\doibase 10.1103/PhysRevE.58.4134} {\bibfield
  {journal} {\bibinfo  {journal} {Phys. Rev. E}\ }\textbf {\bibinfo {volume}
  {58}},\ \bibinfo {pages} {4134} (\bibinfo {year} {1998})}\BibitemShut
  {NoStop}%
\bibitem [{\citenamefont {de~Oliveira}\ \emph {et~al.}(2009)\citenamefont
  {de~Oliveira}, \citenamefont {de~Moura}, \citenamefont {Lyra}, \citenamefont
  {Andrade},\ and\ \citenamefont {Albuquerque}}]{DeOliveira2009}%
  \BibitemOpen
  \bibfield  {author} {\bibinfo {author} {\bibfnamefont {I.~N.}\ \bibnamefont
  {de~Oliveira}}, \bibinfo {author} {\bibfnamefont {F.~A. B.~F.}\ \bibnamefont
  {de~Moura}}, \bibinfo {author} {\bibfnamefont {M.~L.}\ \bibnamefont {Lyra}},
  \bibinfo {author} {\bibfnamefont {J.~S.}\ \bibnamefont {Andrade}}, \ and\
  \bibinfo {author} {\bibfnamefont {E.~L.}\ \bibnamefont {Albuquerque}},\
  }\href {\doibase 10.1103/PhysRevE.79.016104} {\bibfield  {journal} {\bibinfo
  {journal} {Phys. Rev. E}\ }\textbf {\bibinfo {volume} {79}},\ \bibinfo
  {pages} {016104} (\bibinfo {year} {2009})}\BibitemShut {NoStop}%
\bibitem [{\citenamefont {Aydin}\ and\ \citenamefont
  {Sisman}(2014)}]{Aydin2014}%
  \BibitemOpen
  \bibfield  {author} {\bibinfo {author} {\bibfnamefont {A.}~\bibnamefont
  {Aydin}}\ and\ \bibinfo {author} {\bibfnamefont {A.}~\bibnamefont {Sisman}},\
  }\href {\doibase 10.1016/j.physleta.2014.05.044} {\bibfield  {journal}
  {\bibinfo  {journal} {Phys. Lett. Sect. A Gen. At. Solid State Phys.}\
  }\textbf {\bibinfo {volume} {378}},\ \bibinfo {pages} {2001} (\bibinfo {year}
  {2014})},\ \Eprint {http://arxiv.org/abs/1408.1086} {arXiv:1408.1086}
  \BibitemShut {NoStop}%
\bibitem [{\citenamefont {Vidal}\ \emph {et~al.}(2011)\citenamefont {Vidal},
  \citenamefont {Lima},\ and\ \citenamefont {Lyra}}]{Vidal2011}%
  \BibitemOpen
  \bibfield  {author} {\bibinfo {author} {\bibfnamefont {E.~J. G.~G.}\
  \bibnamefont {Vidal}}, \bibinfo {author} {\bibfnamefont {R.~P.~A.}\
  \bibnamefont {Lima}}, \ and\ \bibinfo {author} {\bibfnamefont {M.~L.}\
  \bibnamefont {Lyra}},\ }\href {\doibase 10.1103/PhysRevE.83.061137}
  {\bibfield  {journal} {\bibinfo  {journal} {Phys. Rev. E}\ }\textbf {\bibinfo
  {volume} {83}},\ \bibinfo {pages} {061137} (\bibinfo {year}
  {2011})}\BibitemShut {NoStop}%
\bibitem [{\citenamefont {Ketterle}\ and\ \citenamefont {van
  Druten}(1996)}]{Ketterle1996}%
  \BibitemOpen
  \bibfield  {author} {\bibinfo {author} {\bibfnamefont {W.}~\bibnamefont
  {Ketterle}}\ and\ \bibinfo {author} {\bibfnamefont {N.~J.}\ \bibnamefont {van
  Druten}},\ }\href {\doibase 10.1103/PhysRevA.54.656} {\bibfield  {journal}
  {\bibinfo  {journal} {Phys. Rev. A}\ }\textbf {\bibinfo {volume} {54}},\
  \bibinfo {pages} {656} (\bibinfo {year} {1996})}\BibitemShut {NoStop}%
\bibitem [{\citenamefont {Burioni}\ \emph {et~al.}(2001)\citenamefont
  {Burioni}, \citenamefont {Cassi}, \citenamefont {Rasetti}, \citenamefont
  {Sodano},\ and\ \citenamefont {Vezzani}}]{Burioni2001a}%
  \BibitemOpen
  \bibfield  {author} {\bibinfo {author} {\bibfnamefont {R.}~\bibnamefont
  {Burioni}}, \bibinfo {author} {\bibfnamefont {D.}~\bibnamefont {Cassi}},
  \bibinfo {author} {\bibfnamefont {M.}~\bibnamefont {Rasetti}}, \bibinfo
  {author} {\bibfnamefont {P.}~\bibnamefont {Sodano}}, \ and\ \bibinfo {author}
  {\bibfnamefont {A.}~\bibnamefont {Vezzani}},\ }\href {\doibase
  10.1088/0953-4075/34/23/314} {\bibfield  {journal} {\bibinfo  {journal} {J.
  Phys. B At. Mol. Opt. Phys.}\ }\textbf {\bibinfo {volume} {34}},\ \bibinfo
  {pages} {4697} (\bibinfo {year} {2001})}\BibitemShut {NoStop}%
\bibitem [{\citenamefont {de~Oliveira}\ \emph {et~al.}(2013)\citenamefont
  {de~Oliveira}, \citenamefont {dos Santos}, \citenamefont {de~Moura},
  \citenamefont {Lyra},\ and\ \citenamefont {Serva}}]{DeOliveira2013a}%
  \BibitemOpen
  \bibfield  {author} {\bibinfo {author} {\bibfnamefont {I.~N.}\ \bibnamefont
  {de~Oliveira}}, \bibinfo {author} {\bibfnamefont {T.~B.}\ \bibnamefont {dos
  Santos}}, \bibinfo {author} {\bibfnamefont {F.~A. B.~F.}\ \bibnamefont
  {de~Moura}}, \bibinfo {author} {\bibfnamefont {M.~L.}\ \bibnamefont {Lyra}},
  \ and\ \bibinfo {author} {\bibfnamefont {M.}~\bibnamefont {Serva}},\ }\href
  {\doibase 10.1103/PhysRevE.88.022139} {\bibfield  {journal} {\bibinfo
  {journal} {Phys. Rev. E}\ }\textbf {\bibinfo {volume} {88}},\ \bibinfo
  {pages} {022139} (\bibinfo {year} {2013})}\BibitemShut {NoStop}%
\bibitem [{\citenamefont {Bagnato}\ and\ \citenamefont
  {Kleppner}(1991)}]{Bagnato1991}%
  \BibitemOpen
  \bibfield  {author} {\bibinfo {author} {\bibfnamefont {V.}~\bibnamefont
  {Bagnato}}\ and\ \bibinfo {author} {\bibfnamefont {D.}~\bibnamefont
  {Kleppner}},\ }\href {\doibase 10.1103/PhysRevA.44.7439} {\bibfield
  {journal} {\bibinfo  {journal} {Phys. Rev. A}\ }\textbf {\bibinfo {volume}
  {44}},\ \bibinfo {pages} {7439} (\bibinfo {year} {1991})}\BibitemShut
  {NoStop}%
\bibitem [{\citenamefont {Buonsante}\ \emph {et~al.}(2002)\citenamefont
  {Buonsante}, \citenamefont {Burioni}, \citenamefont {Cassi},\ and\
  \citenamefont {Vezzani}}]{Buonsante2002}%
  \BibitemOpen
  \bibfield  {author} {\bibinfo {author} {\bibfnamefont {P.}~\bibnamefont
  {Buonsante}}, \bibinfo {author} {\bibfnamefont {R.}~\bibnamefont {Burioni}},
  \bibinfo {author} {\bibfnamefont {D.}~\bibnamefont {Cassi}}, \ and\ \bibinfo
  {author} {\bibfnamefont {A.}~\bibnamefont {Vezzani}},\ }\href {\doibase
  10.1103/PhysRevB.66.094207} {\bibfield  {journal} {\bibinfo  {journal} {Phys.
  Rev. B}\ }\textbf {\bibinfo {volume} {66}},\ \bibinfo {pages} {094207}
  (\bibinfo {year} {2002})}\BibitemShut {NoStop}%
\bibitem [{\citenamefont {Serva}(2014)}]{Serva2014a}%
  \BibitemOpen
  \bibfield  {author} {\bibinfo {author} {\bibfnamefont {M.}~\bibnamefont
  {Serva}},\ }\href {\doibase 10.1088/1742-5468/2014/08/P08018} {\bibfield
  {journal} {\bibinfo  {journal} {J. Stat. Mech. Theory Exp.}\ }\textbf
  {\bibinfo {volume} {2014}},\ \bibinfo {pages} {P08018} (\bibinfo {year}
  {2014})}\BibitemShut {NoStop}%
\bibitem [{\citenamefont {Brunelli}\ \emph {et~al.}(2004)\citenamefont
  {Brunelli}, \citenamefont {Giusiano}, \citenamefont {Mancini}, \citenamefont
  {Sodano},\ and\ \citenamefont {Trombettoni}}]{Brunelli2004}%
  \BibitemOpen
  \bibfield  {author} {\bibinfo {author} {\bibfnamefont {I.}~\bibnamefont
  {Brunelli}}, \bibinfo {author} {\bibfnamefont {G.}~\bibnamefont {Giusiano}},
  \bibinfo {author} {\bibfnamefont {F.~P.}\ \bibnamefont {Mancini}}, \bibinfo
  {author} {\bibfnamefont {P.}~\bibnamefont {Sodano}}, \ and\ \bibinfo {author}
  {\bibfnamefont {A.}~\bibnamefont {Trombettoni}},\ }\href {\doibase
  10.1088/0953-4075/37/7/072} {\bibfield  {journal} {\bibinfo  {journal} {J.
  Phys. B At. Mol. Opt. Phys.}\ }\textbf {\bibinfo {volume} {37}},\ \bibinfo
  {pages} {S275} (\bibinfo {year} {2004})}\BibitemShut {NoStop}%
\bibitem [{\citenamefont {Lyra}\ \emph {et~al.}(2014)\citenamefont {Lyra},
  \citenamefont {de~Moura}, \citenamefont {de~Oliveira},\ and\ \citenamefont
  {Serva}}]{Lyra2014a}%
  \BibitemOpen
  \bibfield  {author} {\bibinfo {author} {\bibfnamefont {M.~L.}\ \bibnamefont
  {Lyra}}, \bibinfo {author} {\bibfnamefont {F.~A. B.~F.}\ \bibnamefont
  {de~Moura}}, \bibinfo {author} {\bibfnamefont {I.~N.}\ \bibnamefont
  {de~Oliveira}}, \ and\ \bibinfo {author} {\bibfnamefont {M.}~\bibnamefont
  {Serva}},\ }\href {\doibase 10.1103/PhysRevE.89.052133} {\bibfield  {journal}
  {\bibinfo  {journal} {Phys. Rev. E}\ }\textbf {\bibinfo {volume} {89}},\
  \bibinfo {pages} {052133} (\bibinfo {year} {2014})}\BibitemShut {NoStop}%
\bibitem [{\citenamefont {Cassi}\ and\ \citenamefont
  {Regina}(1993)}]{Cassi1993}%
  \BibitemOpen
  \bibfield  {author} {\bibinfo {author} {\bibfnamefont {D.}~\bibnamefont
  {Cassi}}\ and\ \bibinfo {author} {\bibfnamefont {S.}~\bibnamefont {Regina}},\
  }\href {\doibase 10.1103/PhysRevLett.70.1647} {\bibfield  {journal} {\bibinfo
   {journal} {Phys. Rev. Lett.}\ }\textbf {\bibinfo {volume} {70}},\ \bibinfo
  {pages} {1647} (\bibinfo {year} {1993})}\BibitemShut {NoStop}%
\bibitem [{\citenamefont {Burioni}\ and\ \citenamefont
  {Cassi}(1996)}]{Burioni1996}%
  \BibitemOpen
  \bibfield  {author} {\bibinfo {author} {\bibfnamefont {R.}~\bibnamefont
  {Burioni}}\ and\ \bibinfo {author} {\bibfnamefont {D.}~\bibnamefont
  {Cassi}},\ }\href {\doibase 10.1103/PhysRevLett.76.1091} {\bibfield
  {journal} {\bibinfo  {journal} {Phys. Rev. Lett.}\ }\textbf {\bibinfo
  {volume} {76}},\ \bibinfo {pages} {1091} (\bibinfo {year}
  {1996})}\BibitemShut {NoStop}%
\bibitem [{\citenamefont {Rammal}\ and\ \citenamefont
  {Toulouse}(1983)}]{Rammal1983a}%
  \BibitemOpen
  \bibfield  {author} {\bibinfo {author} {\bibfnamefont {R.}~\bibnamefont
  {Rammal}}\ and\ \bibinfo {author} {\bibfnamefont {G.}~\bibnamefont
  {Toulouse}},\ }\href {\doibase 10.1051/jphyslet:0198300440101300} {\bibfield
  {journal} {\bibinfo  {journal} {J. Phys. Lett.}\ }\textbf {\bibinfo {volume}
  {44}},\ \bibinfo {pages} {13} (\bibinfo {year} {1983})}\BibitemShut {NoStop}%
\bibitem [{\citenamefont {Rammal}(1984)}]{Rammal1984b}%
  \BibitemOpen
  \bibfield  {author} {\bibinfo {author} {\bibfnamefont {R.}~\bibnamefont
  {Rammal}},\ }\href {\doibase 10.1051/jphys:01984004502019100} {\bibfield
  {journal} {\bibinfo  {journal} {J. Phys.}\ }\textbf {\bibinfo {volume}
  {45}},\ \bibinfo {pages} {191} (\bibinfo {year} {1984})}\BibitemShut
  {NoStop}%
\bibitem [{\citenamefont {Alexander}\ and\ \citenamefont
  {Orbach}(1982)}]{Alexander1982a}%
  \BibitemOpen
  \bibfield  {author} {\bibinfo {author} {\bibfnamefont {S.}~\bibnamefont
  {Alexander}}\ and\ \bibinfo {author} {\bibfnamefont {R.}~\bibnamefont
  {Orbach}},\ }\href {\doibase 10.1051/jphyslet:019820043017062500} {\bibfield
  {journal} {\bibinfo  {journal} {J. Phys. Lett.}\ }\textbf {\bibinfo {volume}
  {43}},\ \bibinfo {pages} {625} (\bibinfo {year} {1982})}\BibitemShut
  {NoStop}%
\bibitem [{\citenamefont {Burioni}\ \emph {et~al.}(1999)\citenamefont
  {Burioni}, \citenamefont {Cassi},\ and\ \citenamefont
  {Regina}}]{Burioni1999}%
  \BibitemOpen
  \bibfield  {author} {\bibinfo {author} {\bibfnamefont {R.}~\bibnamefont
  {Burioni}}, \bibinfo {author} {\bibfnamefont {D.}~\bibnamefont {Cassi}}, \
  and\ \bibinfo {author} {\bibfnamefont {S.}~\bibnamefont {Regina}},\ }\href
  {\doibase 10.1016/S0378-4371(98)00477-4} {\bibfield  {journal} {\bibinfo
  {journal} {Phys. A Stat. Mech. its Appl.}\ }\textbf {\bibinfo {volume}
  {265}},\ \bibinfo {pages} {323} (\bibinfo {year} {1999})}\BibitemShut
  {NoStop}%
\bibitem [{\citenamefont {Yang}\ and\ \citenamefont {Zhou}(2012)}]{Yang2012}%
  \BibitemOpen
  \bibfield  {author} {\bibinfo {author} {\bibfnamefont {Z.}~\bibnamefont
  {Yang}}\ and\ \bibinfo {author} {\bibfnamefont {T.}~\bibnamefont {Zhou}},\
  }\href {\doibase 10.1103/PhysRevE.85.056106} {\bibfield  {journal} {\bibinfo
  {journal} {Phys. Rev. E}\ }\textbf {\bibinfo {volume} {85}},\ \bibinfo
  {pages} {056106} (\bibinfo {year} {2012})}\BibitemShut {NoStop}%
\bibitem [{\citenamefont {Chu}\ \emph {et~al.}(2011)\citenamefont {Chu},
  \citenamefont {Zhang}, \citenamefont {Guan},\ and\ \citenamefont
  {Zhou}}]{Chu2011}%
  \BibitemOpen
  \bibfield  {author} {\bibinfo {author} {\bibfnamefont {X.}~\bibnamefont
  {Chu}}, \bibinfo {author} {\bibfnamefont {Z.}~\bibnamefont {Zhang}}, \bibinfo
  {author} {\bibfnamefont {J.}~\bibnamefont {Guan}}, \ and\ \bibinfo {author}
  {\bibfnamefont {S.}~\bibnamefont {Zhou}},\ }\href {\doibase
  10.1016/j.physa.2010.09.038} {\bibfield  {journal} {\bibinfo  {journal}
  {Phys. A Stat. Mech. its Appl.}\ }\textbf {\bibinfo {volume} {390}},\
  \bibinfo {pages} {471} (\bibinfo {year} {2011})}\BibitemShut {NoStop}%
\bibitem [{\citenamefont {Wu}\ \emph {et~al.}(2013)\citenamefont {Wu},
  \citenamefont {Hou},\ and\ \citenamefont {Zhang}}]{Wu2013a}%
  \BibitemOpen
  \bibfield  {author} {\bibinfo {author} {\bibfnamefont {Z.}~\bibnamefont
  {Wu}}, \bibinfo {author} {\bibfnamefont {B.}~\bibnamefont {Hou}}, \ and\
  \bibinfo {author} {\bibfnamefont {H.}~\bibnamefont {Zhang}},\ }\href
  {\doibase 10.1140/epjb/e2013-40246-5} {\bibfield  {journal} {\bibinfo
  {journal} {Eur. Phys. J. B}\ }\textbf {\bibinfo {volume} {86}},\ \bibinfo
  {pages} {405} (\bibinfo {year} {2013})}\BibitemShut {NoStop}%
\bibitem [{\citenamefont {Soh}\ \emph {et~al.}(2010)\citenamefont {Soh},
  \citenamefont {Lim}, \citenamefont {Zhang}, \citenamefont {Fu}, \citenamefont
  {Lee}, \citenamefont {Hung}, \citenamefont {Di}, \citenamefont {Prakasam},\
  and\ \citenamefont {Wong}}]{Soh2010}%
  \BibitemOpen
  \bibfield  {author} {\bibinfo {author} {\bibfnamefont {H.}~\bibnamefont
  {Soh}}, \bibinfo {author} {\bibfnamefont {S.}~\bibnamefont {Lim}}, \bibinfo
  {author} {\bibfnamefont {T.}~\bibnamefont {Zhang}}, \bibinfo {author}
  {\bibfnamefont {X.}~\bibnamefont {Fu}}, \bibinfo {author} {\bibfnamefont
  {G.~K.~K.}\ \bibnamefont {Lee}}, \bibinfo {author} {\bibfnamefont {T.~G.~G.}\
  \bibnamefont {Hung}}, \bibinfo {author} {\bibfnamefont {P.}~\bibnamefont
  {Di}}, \bibinfo {author} {\bibfnamefont {S.}~\bibnamefont {Prakasam}}, \ and\
  \bibinfo {author} {\bibfnamefont {L.}~\bibnamefont {Wong}},\ }\href {\doibase
  10.1016/j.physa.2010.08.015} {\bibfield  {journal} {\bibinfo  {journal}
  {Phys. A Stat. Mech. its Appl.}\ }\textbf {\bibinfo {volume} {389}},\
  \bibinfo {pages} {5852} (\bibinfo {year} {2010})}\BibitemShut {NoStop}%
\bibitem [{\citenamefont {Lu}\ \emph {et~al.}(2006)\citenamefont {Lu},
  \citenamefont {Wang}, \citenamefont {Li},\ and\ \citenamefont
  {Fang}}]{Lu2006}%
  \BibitemOpen
  \bibfield  {author} {\bibinfo {author} {\bibfnamefont {X.~B.}\ \bibnamefont
  {Lu}}, \bibinfo {author} {\bibfnamefont {X.~F.}\ \bibnamefont {Wang}},
  \bibinfo {author} {\bibfnamefont {X.}~\bibnamefont {Li}}, \ and\ \bibinfo
  {author} {\bibfnamefont {J.~Q.}\ \bibnamefont {Fang}},\ }\href {\doibase
  10.1016/j.physa.2006.02.037} {\bibfield  {journal} {\bibinfo  {journal}
  {Phys. A Stat. Mech. its Appl.}\ }\textbf {\bibinfo {volume} {370}},\
  \bibinfo {pages} {381} (\bibinfo {year} {2006})}\BibitemShut {NoStop}%
\bibitem [{\citenamefont {Huang}(2006)}]{Huang2006}%
  \BibitemOpen
  \bibfield  {author} {\bibinfo {author} {\bibfnamefont {D.}~\bibnamefont
  {Huang}},\ }\href {\doibase 10.1103/PhysRevE.74.046208} {\bibfield  {journal}
  {\bibinfo  {journal} {Phys. Rev. E}\ }\textbf {\bibinfo {volume} {74}},\
  \bibinfo {pages} {046208} (\bibinfo {year} {2006})}\BibitemShut {NoStop}%
\bibitem [{\citenamefont {Lin}\ and\ \citenamefont {Zhang}(2013)}]{Lin2013b}%
  \BibitemOpen
  \bibfield  {author} {\bibinfo {author} {\bibfnamefont {Y.}~\bibnamefont
  {Lin}}\ and\ \bibinfo {author} {\bibfnamefont {Z.}~\bibnamefont {Zhang}},\
  }\href {\doibase 10.1103/PhysRevE.87.062140} {\bibfield  {journal} {\bibinfo
  {journal} {Phys. Rev. E}\ }\textbf {\bibinfo {volume} {87}},\ \bibinfo
  {pages} {1} (\bibinfo {year} {2013})},\ \Eprint
  {http://arxiv.org/abs/arXiv:1307.0903v1} {arXiv:arXiv:1307.0903v1}
  \BibitemShut {NoStop}%
\bibitem [{\citenamefont {Zhang}\ \emph {et~al.}(2013)\citenamefont {Zhang},
  \citenamefont {Shan},\ and\ \citenamefont {Chen}}]{Zhang2013}%
  \BibitemOpen
  \bibfield  {author} {\bibinfo {author} {\bibfnamefont {Z.}~\bibnamefont
  {Zhang}}, \bibinfo {author} {\bibfnamefont {T.}~\bibnamefont {Shan}}, \ and\
  \bibinfo {author} {\bibfnamefont {G.}~\bibnamefont {Chen}},\ }\href {\doibase
  10.1103/PhysRevE.87.012112} {\bibfield  {journal} {\bibinfo  {journal} {Phys.
  Rev. E}\ }\textbf {\bibinfo {volume} {87}},\ \bibinfo {pages} {012112}
  (\bibinfo {year} {2013})},\ \Eprint {http://arxiv.org/abs/arXiv:1212.5998v1}
  {arXiv:arXiv:1212.5998v1} \BibitemShut {NoStop}%
\bibitem [{\citenamefont {Baronchelli}\ and\ \citenamefont
  {Pastor-Satorras}(2010)}]{Baronchelli2010}%
  \BibitemOpen
  \bibfield  {author} {\bibinfo {author} {\bibfnamefont {A.}~\bibnamefont
  {Baronchelli}}\ and\ \bibinfo {author} {\bibfnamefont {R.}~\bibnamefont
  {Pastor-Satorras}},\ }\href {\doibase 10.1103/PhysRevE.82.011111} {\bibfield
  {journal} {\bibinfo  {journal} {Phys. Rev. E}\ }\textbf {\bibinfo {volume}
  {82}},\ \bibinfo {pages} {011111} (\bibinfo {year} {2010})}\BibitemShut
  {NoStop}%
\bibitem [{\citenamefont {Baronchelli}\ \emph {et~al.}(2011)\citenamefont
  {Baronchelli}, \citenamefont {Castellano},\ and\ \citenamefont
  {Pastor-Satorras}}]{Baronchelli2011}%
  \BibitemOpen
  \bibfield  {author} {\bibinfo {author} {\bibfnamefont {A.}~\bibnamefont
  {Baronchelli}}, \bibinfo {author} {\bibfnamefont {C.}~\bibnamefont
  {Castellano}}, \ and\ \bibinfo {author} {\bibfnamefont {R.}~\bibnamefont
  {Pastor-Satorras}},\ }\href {\doibase 10.1103/PhysRevE.83.066117} {\bibfield
  {journal} {\bibinfo  {journal} {Phys. Rev. E}\ }\textbf {\bibinfo {volume}
  {83}},\ \bibinfo {pages} {066117} (\bibinfo {year} {2011})}\BibitemShut
  {NoStop}%
\bibitem [{\citenamefont {Suchecki}\ \emph {et~al.}(2005)\citenamefont
  {Suchecki}, \citenamefont {Egu{\'{i}}luz},\ and\ \citenamefont {{San
  Miguel}}}]{Suchecki2005}%
  \BibitemOpen
  \bibfield  {author} {\bibinfo {author} {\bibfnamefont {K.}~\bibnamefont
  {Suchecki}}, \bibinfo {author} {\bibfnamefont {V.~M.}\ \bibnamefont
  {Egu{\'{i}}luz}}, \ and\ \bibinfo {author} {\bibfnamefont {M.}~\bibnamefont
  {{San Miguel}}},\ }\href {\doibase 10.1103/PhysRevE.72.036132} {\bibfield
  {journal} {\bibinfo  {journal} {Phys. Rev. E}\ }\textbf {\bibinfo {volume}
  {72}},\ \bibinfo {pages} {036132} (\bibinfo {year} {2005})}\BibitemShut
  {NoStop}%
\bibitem [{\citenamefont {Chen}\ and\ \citenamefont {Zhang}(2007)}]{Chen2007}%
  \BibitemOpen
  \bibfield  {author} {\bibinfo {author} {\bibfnamefont {H.}~\bibnamefont
  {Chen}}\ and\ \bibinfo {author} {\bibfnamefont {F.}~\bibnamefont {Zhang}},\
  }\href {\doibase 10.1016/j.dam.2006.09.008} {\bibfield  {journal} {\bibinfo
  {journal} {Discret. Appl. Math.}\ }\textbf {\bibinfo {volume} {155}},\
  \bibinfo {pages} {654} (\bibinfo {year} {2007})}\BibitemShut {NoStop}%
\bibitem [{\citenamefont {Teplyaev}(1998)}]{Teplyaev1998a}%
  \BibitemOpen
  \bibfield  {author} {\bibinfo {author} {\bibfnamefont {A.}~\bibnamefont
  {Teplyaev}},\ }\href {\doibase 10.1006/jfan.1998.3297} {\bibfield  {journal}
  {\bibinfo  {journal} {J. Funct. Anal.}\ }\textbf {\bibinfo {volume} {159}},\
  \bibinfo {pages} {537} (\bibinfo {year} {1998})}\BibitemShut {NoStop}%
\bibitem [{\citenamefont {Hare}\ \emph {et~al.}(2012)\citenamefont {Hare},
  \citenamefont {Steinhurst}, \citenamefont {Teplyaev},\ and\ \citenamefont
  {Zhou}}]{Hare2012}%
  \BibitemOpen
  \bibfield  {author} {\bibinfo {author} {\bibfnamefont {K.~E.}\ \bibnamefont
  {Hare}}, \bibinfo {author} {\bibfnamefont {B.~A.}\ \bibnamefont
  {Steinhurst}}, \bibinfo {author} {\bibfnamefont {A.}~\bibnamefont
  {Teplyaev}}, \ and\ \bibinfo {author} {\bibfnamefont {D.}~\bibnamefont
  {Zhou}},\ }\href {\doibase 10.4310/MRL.2012.v19.n3.a3} {\bibfield  {journal}
  {\bibinfo  {journal} {Math. Res. Lett.}\ }\textbf {\bibinfo {volume} {19}},\
  \bibinfo {pages} {537} (\bibinfo {year} {2012})}\BibitemShut {NoStop}%
\bibitem [{\citenamefont {Tsallis}\ \emph {et~al.}(1997)\citenamefont
  {Tsallis}, \citenamefont {da~Silva}, \citenamefont {Mendes}, \citenamefont
  {Vallejos},\ and\ \citenamefont {Mariz}}]{Tsallis1997}%
  \BibitemOpen
  \bibfield  {author} {\bibinfo {author} {\bibfnamefont {C.}~\bibnamefont
  {Tsallis}}, \bibinfo {author} {\bibfnamefont {L.~R.}\ \bibnamefont
  {da~Silva}}, \bibinfo {author} {\bibfnamefont {R.~S.}\ \bibnamefont
  {Mendes}}, \bibinfo {author} {\bibfnamefont {R.~O.}\ \bibnamefont
  {Vallejos}}, \ and\ \bibinfo {author} {\bibfnamefont {A.~M.}\ \bibnamefont
  {Mariz}},\ }\href {\doibase 10.1103/PhysRevE.56.R4922} {\bibfield  {journal}
  {\bibinfo  {journal} {Phys. Rev. E}\ }\textbf {\bibinfo {volume} {56}},\
  \bibinfo {pages} {R4922} (\bibinfo {year} {1997})}\BibitemShut {NoStop}%
\bibitem [{\citenamefont {Xie}\ \emph {et~al.}(2015)\citenamefont {Xie},
  \citenamefont {Lin},\ and\ \citenamefont {Zhang}}]{Xie2015}%
  \BibitemOpen
  \bibfield  {author} {\bibinfo {author} {\bibfnamefont {P.}~\bibnamefont
  {Xie}}, \bibinfo {author} {\bibfnamefont {Y.}~\bibnamefont {Lin}}, \ and\
  \bibinfo {author} {\bibfnamefont {Z.}~\bibnamefont {Zhang}},\ }\href
  {\doibase 10.1063/1.4922265} {\bibfield  {journal} {\bibinfo  {journal} {J.
  Chem. Phys.}\ }\textbf {\bibinfo {volume} {142}} (\bibinfo {year} {2015}),\
  10.1063/1.4922265}\BibitemShut {NoStop}%
\bibitem [{\citenamefont {Xie}\ \emph {et~al.}(2016)\citenamefont {Xie},
  \citenamefont {Zhang},\ and\ \citenamefont {Comellas}}]{Xie2016a}%
  \BibitemOpen
  \bibfield  {author} {\bibinfo {author} {\bibfnamefont {P.}~\bibnamefont
  {Xie}}, \bibinfo {author} {\bibfnamefont {Z.}~\bibnamefont {Zhang}}, \ and\
  \bibinfo {author} {\bibfnamefont {F.}~\bibnamefont {Comellas}},\ }\href
  {\doibase 10.1016/j.amc.2016.04.033} {\bibfield  {journal} {\bibinfo
  {journal} {Appl. Math. Comput.}\ }\textbf {\bibinfo {volume} {286}},\
  \bibinfo {pages} {250} (\bibinfo {year} {2016})}\BibitemShut {NoStop}%
\end{thebibliography}%

\end{document}